# Magnetic Excitations of Spin-Gap System $Na_3Cu_2SbO_6$ with Distorted Honeycomb Structure

Yoko Miura, Yukio Yasui, Taketo Moyoshi, Masatoshi Sato* and Kazuhisa Kakurai[1]

*Department of Physics, Nagoya University, Furo-cho, Chikusa-ku, Nagoya 464-8602*
[1]*Quantum Beam Science Directorate, Japan Atomic Energy Agency,
Tokai-mura, Naka-gun, Ibaraki 319-1195.*

**Abstract**

Magnetic excitation spectra of quantum spins of $Na_3Cu_2SbO_6$ with the distorted honeycomb structure have been measured by neutron inelastic scattering. The intensity distribution and the dispersion curves of the excitations indicate that the spin system can be well understood by considering one-dimensional spin chains formed of $Cu^{2+}$ ions on a lattice of the alternation of the shorter- and longer-spacings along the $b$ axis. The exchange interactions between the neighboring spins with the shorter- and longer-spacings correspond to ferromagnetic and antiferromagnetic ones, respectively. These findings are consistent with those we reported previously on the basis of macroscopic measurements. The observed gap $\Delta_G$ is 8.9 meV and the antiferromagnetic- and ferromagnetic-interactions have been found to be 13.9 and -12.5 meV, respectively.

*Corresponding author: e43247a@nucc.cc.nagoya-u.ac.jp

## 1. Introduction

$Na_3Cu_2SbO_6$ has a honeycomb lattice of edge-sharing $CuO_6$ octahedra,[1] but it is distorted due to the $Cu^{2+}$ Jahn-Teller effect. The valence of Cu atoms is +2 and the spin $S$ is 1/2. For this quantum spins, the existence of the excitation gap (spin gap) was found by the study of macroscopic quantities.[2] Figure 1 schematically shows the $Cu^{2+}$ skeleton lattice viewed from the direction perpendicular to the $ab$-plane. The ratio of the lattice parameter $b$ to $a$ is smaller than $\sqrt{3}$.

We define the superexchange interactions between two Cu spins, $J_1$, $J_1'$ and $J_2$ as shown in Fig. 1, then, neglect $J_1'$, considering the rather large Jahn-Teller distortion working on the $Cu^{2+}$ spins in the $x^2-y^2$ orbit (with local coordinates),[2] and obtain the alternating chain model along the $b$ axis with $J_1$ and $J_2$. Actually, the antiferromagnetic(AF)-ferromagnetic(F) alternating chain model was found, as reported in ref. 2, to successfully describe the observed magnetic and thermal behaviors, where the gap energy $\Delta_G$ of ~90 K, and the F and AF exchange interactions $J_F$ and $J_{AF}$ of -209 K and 165 K were obtained, respectively, by using the Hamiltonian $H = \Sigma_i\{J_1 S_{2i}\cdot S_{2i+1}+J_2 S_{2i+1}\cdot S_{2i+2}\}$. However, we could not distinguish which one of $J_1$ and $J_2$ is F or AF. Similar studies on the spin gap phenomenon has also been carried out for $Na_2Cu_2TeO_6$ by Xu et al.[3]

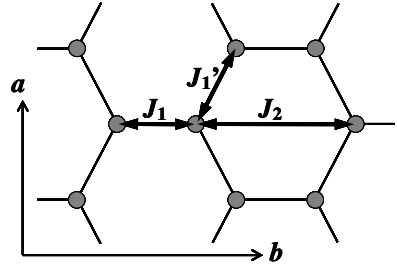

**Fig. 1** Schematic figure of $Cu^{2+}$ honeycomb lattice of $Na_3Cu_2SbO_6$ viewed from the direction perpendicular to the $ab$-plane, where only the Cu sites are shown. The exchange interactions $J_1$, $J_1'$ and $J_2$ are defined between the $Cu^{2+}$ spins.

Deakhshan et al.[4] reported by their theoretical study that the AF-AF alternating chain model is appropriate for $Na_3Cu_2SbO_6$. Contrary to their result, Koo and Whangbo[5] have reported that the AF-F alternating chain model is appropriate to the description of the magnetic properties of the system based on their band calculations.

The present neutron scattering study has been carried out to determine the values of $J_1$ and $J_2$ by using single crystals of $Na_3Cu_2SbO_6$, and here we present detailed data, which clearly show that the system has the AF-F alternating chains with ferromagnetic $J_1$.

## 2. Experiments

Single crystals of $Na_3Cu_2SbO_6$ were prepared from mixtures of CuO, $Sb_2O_3$ and $Na_2CO_3$ with a condition of slightly excess Na. They were ground and heated up rapidly to 1150 °C in a crucible, and kept at the temperature for 2 h, cooled to about 1000 °C at a rate of ~2 K/h and furnace cooled to room temperature. The neutron measurements were carried out on the triplet-axis-spectrometer ISSP-PONTA at JRR-3 of Japan Atomic Energy Agency in Tokai. Three crystals aligned within 0.7 ° were used, where the typical volume of each crystal was about 60 mm$^3$. The 002 reflection of the pyrolytic graphite (PG) was used both for the monochromator and the analyzer. A PG filter after the sample was used to suppress the higher order contaminations. Horizontal collimations were effectively 40'-40'-80'-80'. The sample crystals were oriented with the [100] and [010] axes in the scattering plane. The condition of the fixed final neutron energy $E_f = 14.7$ meV was adopted.

## 3. Experimental Results and Discussion

Because the energy gap $\Delta_G$ of the singlet-triplet excitation is, as described later, were found to be about 9 meV, we collected the scattering intensities $I$ at the transfer energy ($\Delta E$) of 9 meV to find the positions of the scattering vector ($Q$) corresponding to the minimum energy of the excitation. It was carried out at 10 K by scanning $k$ along $(h, k, 0)$ for several fixed values of $h$ and the results are shown in Fig. 2(a). We can see, for $h = 1$, the intensity peaks of the singlet-triplet excitation at the points $k = 0.5, 2.5$ and 3.5 in the region of $k < 4.0$. For $h = 1.5$ and 2, peaks are found at the same $k$ points as for $h = 1$. We do not see a peak in the $k$-scans along $(1, k, 0)$, $(1.5, k, 0)$ and $(2, k, 0)$ at $k = 1.0$, and there is no peak in the data taken by the $h$-scan along $(h, 1, 0)$, either (not shown). These results indicate that the scattering intensities along $(h, 1, 0)$ correspond to the background (BG). By considering this fact, the intrinsic part of the magnetic scattering intensities at $(h, 2.5, 0)$ were determined by subtracting the neutron counts at $(h, 1, 0)$ from those at $(h, 2.5, 0)$, and the result is shown in Fig. 2(b) together with the $h$-dependence (solid line) calculated by considering the form factor of the Cu $3d$ $x^2$-$y^2$ spins (with local coordinates) and the one-dimensional nature of the spin chains along $b$. From the figure, the one-dimensionality is found to describe the observed $h$ dependence well.

From the data shown in Fig. 2(a), we have found that the energy minima of the singlet-triplet excitation are found on the streaks along $(h, 0.5, 0)$, $(h, 2.5, 0)$ and $(h, 3.5, 0)$ in the reciprocal space. This type of spectral distribution in the $Q$ space corresponds to the spin correlation shown in Fig. 2(c), although the spins of the system do not have the static correlation. If the static correlation existed in the chain, we had magnetic Bragg streaks along $(h, 0.5, 0)$, $(h, 2.5, 0)$ and $(h, 3.5, 0)$. Similar data can be found in the data reported for the F-AF alternating spin chains of $CuNb_2O_6$,[6] for example. From the result, $J_1$ and $J_2$ are found to be F($J_1<0$) and AF($J_2>0$), respectively.

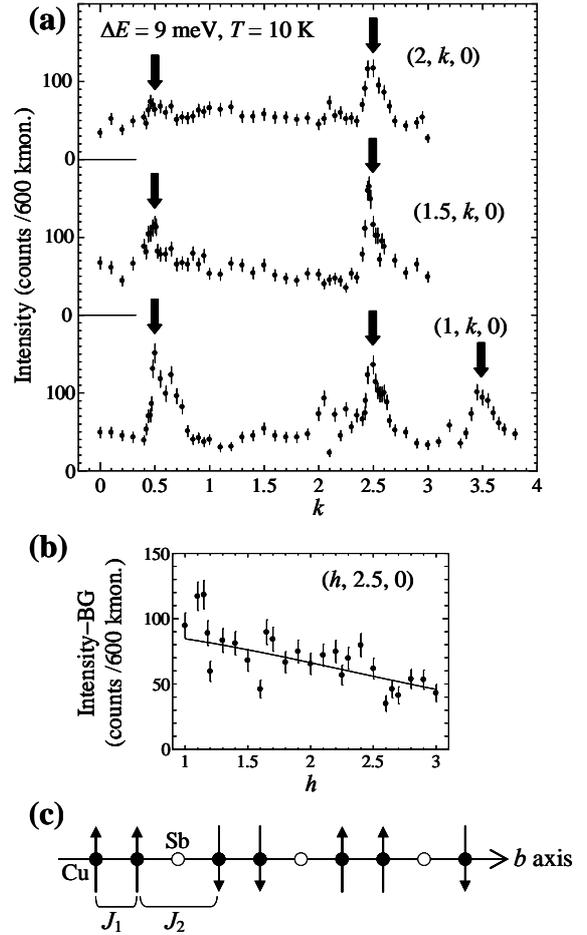

**Fig. 2** Results of the neutron scattering measurements carried out at $\Delta E = 9$ meV and $T = 10$ K. (a) Data of the $k$-scan at several fixed $h$ values along $(h, k, 0)$. (b) Data of $h$-scan at $Q = (h, 2.5, 0)$ after subtracting the background(BG). The solid line shows the $h$-dependence calculated by considering the form factor of the Cu $3d$ $x^2$-$y^2$ spins (with local coordinates) for the chains with one dimensional nature. (c) Schematic figure of the spin correlation which corresponds to that of the present system. The solid and open circles are Cu and Sb atoms, respectively. The arrows show the spin directions. Details are in the text.

In Fig. 3, the BG-subtracted intensities $\{I(Q, \Delta E) - BG\}$ of the magnetic excitation at $Q = (1, 2.5, 0)$ are plotted against $\Delta E$ at several fixed temperatures $T$, where BG is assumed to be independent of $\Delta E$ and $T$.

We can see that the intensity-$\Delta E$ curve in the low energy region begins to become concave with decreasing $T$ at around $T_0$, where the magnetic susceptibility has a maximum value.

In the analyses of the data in Fig. 3, we express the magnetic scattering intensity $S(q, \omega)$ as

$$S(q, \omega) \sim \chi(q) \cdot \Gamma \omega \cdot \{1-\exp(-\hbar\omega/k_B T)\}^{-1} \cdot \{1/[(\omega-\omega_q)^2+\Gamma^2]+1/[(\omega+\omega_q)^2+\Gamma^2]\}, \quad (1)$$

where $\Gamma$ and $\omega_q$ are the half width at half maximum and energy of the triplet excitation with the wave vector $q$, respectively. Here, $\omega_q$ can be approximated to be $q$ independent within the ellipsoid of the $q$ resolution centered at $Q = (1, 2.5, 0)$ corresponding to the dispersion bottom. After convoluting the calculated $S(q, \omega)$ with the resolution function $R(q-Q, \omega-\Delta E)$, $S(Q, \Delta E)$ has been obtained and fitted to the data shown in Fig. 3, and the results are presented by the solid lines.

The values of $\Gamma$ and $\omega_Q$ (or the value of $\Delta_G$) obtained in the above analyses are plotted against $T$ in the upper and lower insets of Fig. 3, respectively. From these figures, we find that $\Delta_G$ does not decrease with increasing $T$, and the concave structure observed in Fig. 3 can be understood as the result of the decrease of $\Gamma$ with decreasing $T$, and the apparent gap grows as $T$ approaches zero.

To study the dispersion relation of the excitations at low temperature, the scattering profiles were measured at 10 K by scanning $\Delta E$ at $(1, 2.5, 0)$ and at several points of $(1.5, k, 0)$ in the region of $2.4 \geq k > 2$, and their examples are shown in Fig. 4. The observed profiles have been fitted by the calculated ones obtained after convoluting $S(q, \omega)$ with the resolution function $R(q-Q, \omega-\Delta E)$, and the results are shown in the figure by the solid lines. In the fittings, we have used the simple expression $S(q, \omega) \sim \chi(q) \cdot \{1-\exp(-\hbar\omega/k_B T)\}^{-1} \cdot \{\Gamma/[(\omega-\omega_Q)^2+\Gamma^2]\}$ instead of eq. (1), because $\Gamma$ is much smaller than $\omega_q$ at this low temperature. The dispersion curve or the $k$ dependence of $\omega_q$ used in the calculation was determined self-consistently, and it is shown in Fig. 5(a). We have also used an assumption that BG is linear in $\Delta E$. The gap energy $\Delta_G$ is found to be ~ 8.9 meV. Figure 5(b) shows the scale factors $\chi(Q)$, where the form-factor-correction was made. In the figure, we find an interesting characteristic of the excitation that the peak is at the incommensurate $Q$ point with $k$ ~ 2.4. In Fig. 5(c), $\Gamma$ is plotted against $k$. Although it seems to exhibit slight $k$ dependence, we do not know what it indicates at this moment.

To find information on the physical parameters of the present system, it is useful to compare the above data with results of model calculations by Watanabe and Yokoyama[7] reported for alternating spin chains. For example, the dispersion curve in Fig. 5(a) can be well explained by choosing the parameters $J_{AF}$ ~ 13.9 meV and $J_F/J_{AF}$ ~ -0.9, which should be compared with values of $J_{AF}$ ~ 14.2 meV and $J_F/J_{AF}$ ~ -1.27, obtained previously from the macroscopic measurements.[2] The $\Delta_G$ value of 8.9 meV should also be compared with the previous value of ~ 7.8 meV. The exchange parameters obtained here are qualitatively different from the prediction of ref. 4 that both $J_1$ and $J_2$ are AF. The calculated values of ref. 5 agree qualitatively with ours. However, their $J_F/J_{AF}$ value exhibits rather large difference from the present one.

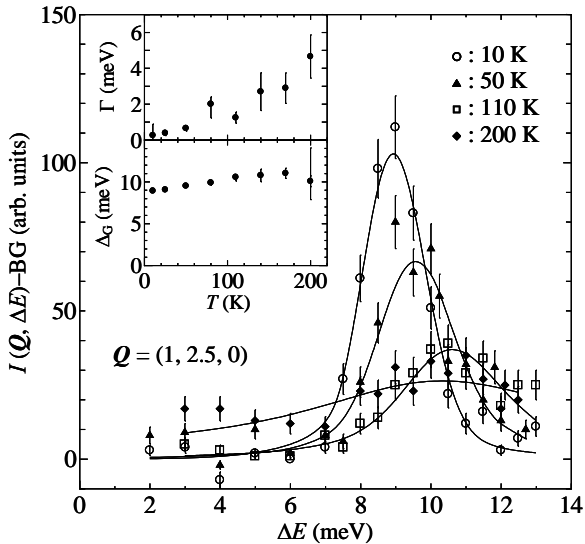

**Fig. 3** BG-subtracted intensities $\{I(Q, \Delta E)-BG\}$ of the magnetic excitation are plotted for $Q = (1, 2.5, 0)$ against $\Delta E$ at several fixed temperatures. The solid lines are the fitted curves. The upper and lower insets show the values of $\Gamma$ and $\Delta_G$ against $T$ obtained in the present analyses. Details are in the text.

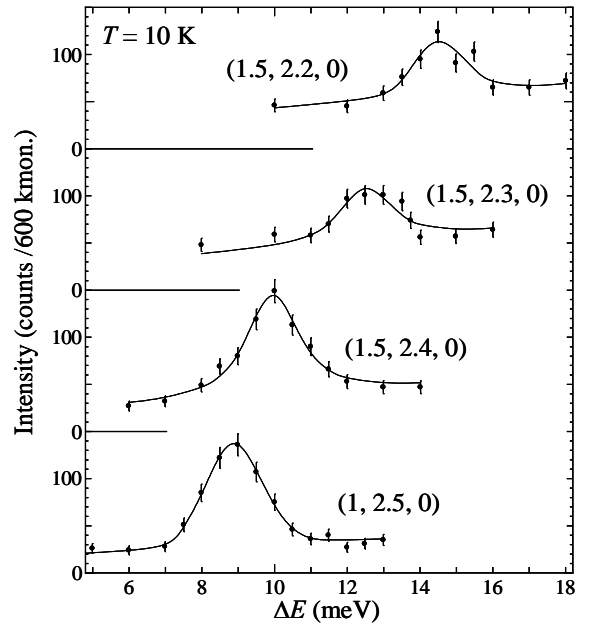

**Fig. 4** Typical profiles of the $\Delta E$-scans at various fixed $Q$-points. The solid lines are the results of the fittings obtained by using Lorentzian type broadening of the excitations.

As stated above, the scattering intensity has a peak at the incommensurate $Q$ position with $k \sim 2.4$. It is consistent with the reported results of ref. 7, that is, for the spin system on the lattice with the alteration of the shorter- and longer-spacings and $J_1/J_2 \sim -0.9$, the shift toward the smaller $k$ side from $k = 2.5$ can be deduced from the data shown in their Fig. 10.

## 4. Conclusions

Magnetic excitation spectra of the spins of $Na_3Cu_2SbO_6$ have been measured by neutron inelastic scattering. The data show that the spin system can be well understood by the one-dimensional spin chains formed of $Cu^{2+}$ ions on a lattice of the alternation of the shorter- and longer-spacings along the $b$ axis. The exchange interaction between the neighboring spins with the shorter distance ($J_1$) is found to be ferromagnetic and that with the longer spacing ($J_2$) is antiferromagnetic. The observed gap $\Delta_G$ is 8.9 meV and the antiferromagnetic- and ferromagnetic-interactions have been found to be 13.9 and -12.5 meV, respectively. These values are basically consistent with those we reported previously on the basis of macroscopic measurements. The magnetic scattering intensity was found to have a peak at the incommensurate positions in the reciprocal space, which can also be understood by using the calculated results by Watanabe and Yokoyama.[7]

**Acknowledgements**

The authors thank for Prof. Y. Ono of Niigata University for his valuable discussion. Neutron scattering at JRR-3 was performed within the national user's program of the Neutron Science Laboratory of the Institute for Solid State Physics (NSL-ISSP). This work is supported by Grants-in-Aid for Scientific Research from the Japan Society for the Promotion of Science (JSPS) and by Grants-in-Aid on Priority Areas from the Ministry of Education, Culture, Sports, Science and Technology.

**References**
1) O. A. Smirnova, V. B. Nalbandyan, A. A. Petrenko and M. Avdeev: J. Solid State Chem. 178 (2005) 1165.
2) Y. Miura, R. Hirai, Y. Kobayashi and M. Sato: J. Phys. Soc. Jpn. 75 (2006) 084707.
3) J. Xu, A. Assoud, N. Soheilnia, S. Derakhshan, H. L. Cuthbert, J. E. Greedan, M. H. Whangbo and H. Kleinke: Inorg. Chem. 44 (2005) 5042.
4) S. Derakhshan, H. L. Cuthbert, J. E. Greedan, B. Rahaman and T. Saha-Dasgupta: Phys. Rev. B 76 (2007) 104403.
5) H. -J. Koo and M. -H. Whangbo: Inorg. Chem. 47 (2008) 128.
6) K. Kodama, H. Harashina, H. Sasaki, M. Kato, M. Sato, K. Kakurai and M. Nishi: J. Phys. Soc. Jpn. 68 (1999) 237.
7) S. Watanabe and H. Yokoyama: J. Phys. Soc. Jpn. 68 (1999) 2073.

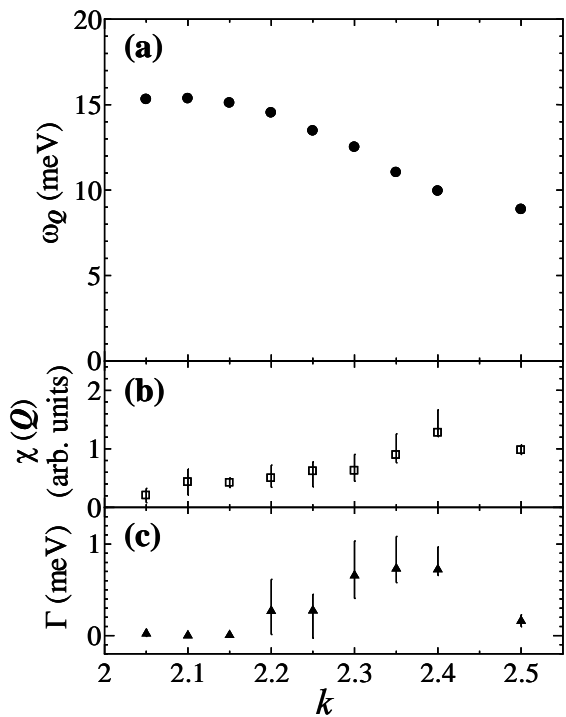

**Fig. 5** The dispersion curve along $b^*$ (a), the scale factor $\chi(Q)$ corrected by the form factor (b), and the $\Gamma$ value (c) of the magnetic excitation are shown against $k$.

4